\documentclass{iaus}
\usepackage{graphicx}

\title[VLBI multi-epoch water maser observations toward massive protostars] 
{VLBI multi-epoch water maser observations toward massive protostars}

\author[J. M. Torrelles, J. F. G\'omez, N. A. Patel, S. Curiel, G. Anglada, \& R. Estalella]   
{Jos\'e M. Torrelles$^1$, Jos\'e F. G\'omez$^2$, Nimesh A. Patel$^3$,\\ Salvador Curiel$^4$, Guillem Anglada$^2$, \and Robert Estalella$^5$}

\affiliation{$^1$ICE(CSIC)-UB/IEEC, Barcelona (Spain)\\[\affilskip]
$^2$IAA(CSIC), Granada (Spain)\\[\affilskip]
$^3$Harvard-Smithsonian, CfA, Cambridge (USA)\\[\affilskip]
$^4$IAUNAM, M\'exico D.F. (M\'exico)\\[\affilskip]
$^5$UB/IEEC, Barcelona (Spain)}

\pubyear{2012}
\volume{287}  
\pagerange{119--126}
\jname{Cosmic Masers - from OH to H$_0$}
\editors{R. Booth, L. Humphries \& W. Vlemmings, eds.}
\begin{document}

\maketitle

\begin{abstract}
VLBI multi-epoch water maser observations are a powerful tool to
study the gas very close to the central engine responsible for the phenomena
associated with the early evolution of massive protostars. In this paper we present a summary
of the main observational results obtained toward the massive
star-forming 
regions of Cepheus
A and W75N.  These
observations revealed unexpected phenomena in the earliest stages
of  evolution of massive objects (e.g., non-collimated ``short-lived'' pulsed
ejections in different massive protostars), and provided new insights in the
study of the dynamic scenario of the formation of high-mass stars
(e.g., simultaneous presence of a jet and wide-angle outflow in the
massive object Cep A HW2, similar to what is observed  in low-mass
protostars). In addition, with these observations it has been possible to identify new, previously unseen centers of high-mass
star formation through outflow activity.

\end{abstract}

\firstsection 
\section{Introduction}

It is well-established that low-mass stars form via an accretion
process. A natural consequence of this process is the formation,
during the early stages of evolution of a young stellar object (YSO),  
of a system (with typical scales of
$\simeq 100$ AU) that comprises a central
protostar, surrounded by a circumstellar (protoplanetary) disk, and a
collimated outflow, ejected perpendicular to the disk (e.g., Anglada 1996). The disk is 
the reservoir of material from which the
central protostar accretes further matter, while the collimated outflow
releases the necessary angular momentum and magnetic flux for this
accretion to proceed. This accretion scenario seems to be generally
applicable in the formation of stars up to $\simeq 20$ M$_{\odot}$
(e.g., Garay \& Lizano 1999). However, very few massive
protostar-disk-outflow systems have been identified and studied in
detail, at scales $\leq 3000$ AU (Patel et al. 2005, Jim\'enez-Serra
et al. 2007, Torrelles et al. 2007,  Zapata et al. 2009, Davies et
al. 2010, Carrasco-Gonz\'alez et al. 2010a, 2011, Fern\'andez-L\'opez
et al. 2011). This scarcity of studies 
is probably due to observational limitations (sensitivity
and angular resolution), given that high-mass stars are more rare, are typically located at
larger distances, and form in a more clustered environment than their
low-mass counterparts. 

Observations of water maser emission  ($\lambda = 1.35$ cm) towards massive
YSOs can overcome these observational
limitations (sensitivity and angular resolution). Water maser emission 
 is compact ($\leq 1$ mas) and
strong (brightness temperatures can reach $10^{10}$ K), which make it
an ideal tool for observations with Very Long Baseline
Interferometry (VLBI), i.e., with angular resolutions $< 1$
mas. Therefore, we can study shocked, warm ($\sim 500$ K), and dense
($\simeq$ $10^8$-$10^9$ cm$^{-3}$) gas, at scales of only 10-1000 AU
from massive protostars (Reid \& Moran 1981), with the possibility
of accurately tracing their kinematics, by measuring proper motions of
a few km s$^{-1}$ in timescales of a few weeks. In this paper we
summarize the main results of our studies obtained toward the high-mass star forming
regions of Cepheus A and W75N, 
using VLBI multi-epoch water maser
observations with an angular resolution of $\sim$ 0.5~mas. These results were reported by
Torrelles et al. (2001a,b, 2003, 2011) and summarized in Torrelles et al. (2012). 
Other very interesting examples (not included in this paper) can be found in the studies on IRAS 06061+2151 (Motogi et al. 2008), AFGL 5142 (Goddi, Moscadelli, \& Sanna 2011), AFGL 2591 (Sanna et al. 2012, Trinidad et al. 2003), and Cep A HW3d (Chibueze et al. 2012) (see also talks in this conference by A. Bartkiewicz-H. van Langevelde, J. Chibueze, C. Goddi, T. Hirota, and A. Sanna).

\begin{figure}[!ht]
\begin{center} 
\includegraphics[scale=0.6]{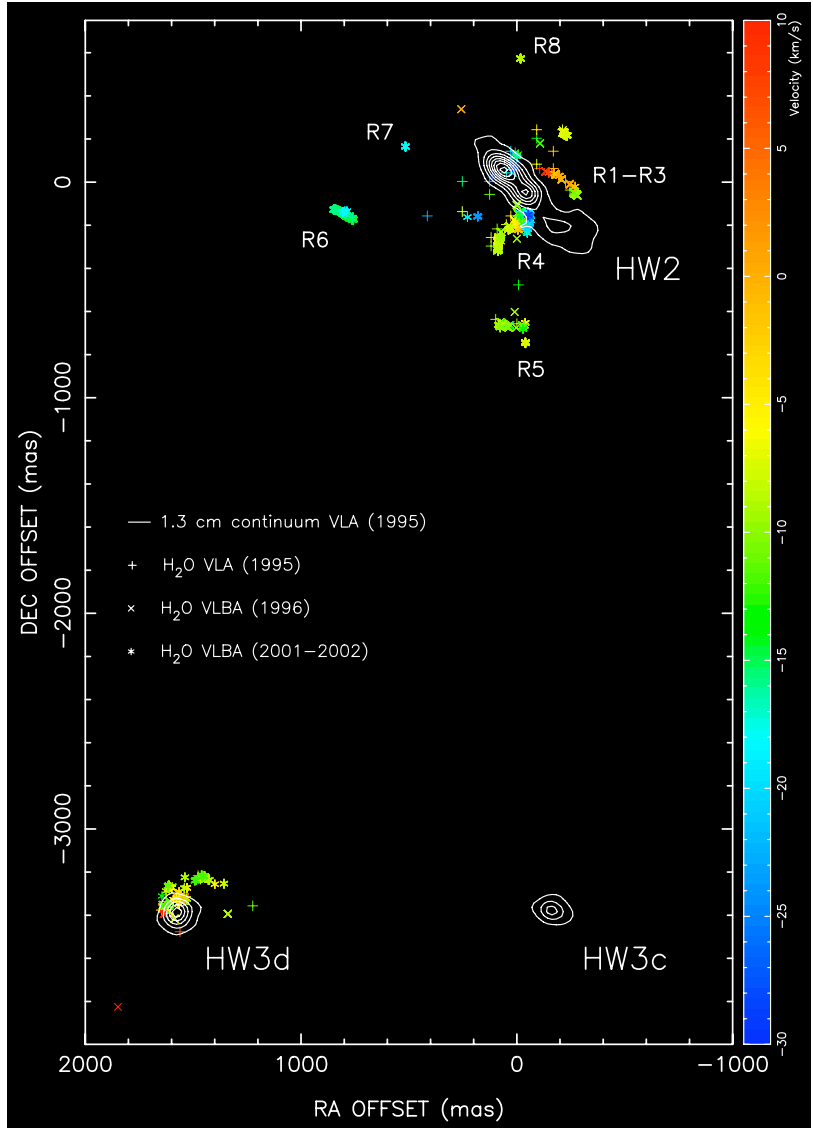} 
\end{center} 
\caption{(Published online) Positions and radial velocities (color code) of the H$_2$O
  masers overlaid onto the 1.3~cm continuum maps (contours) of Cep A HW2,
  HW3c, and HW3d. Sub-regions ``R'' discussed in this paper are
  numbered. (Figure from Torrelles et al. 2011).}
\end{figure}
\section{Cepheus A}

\begin{figure}[!ht]
\begin{center} \includegraphics[scale=0.6]{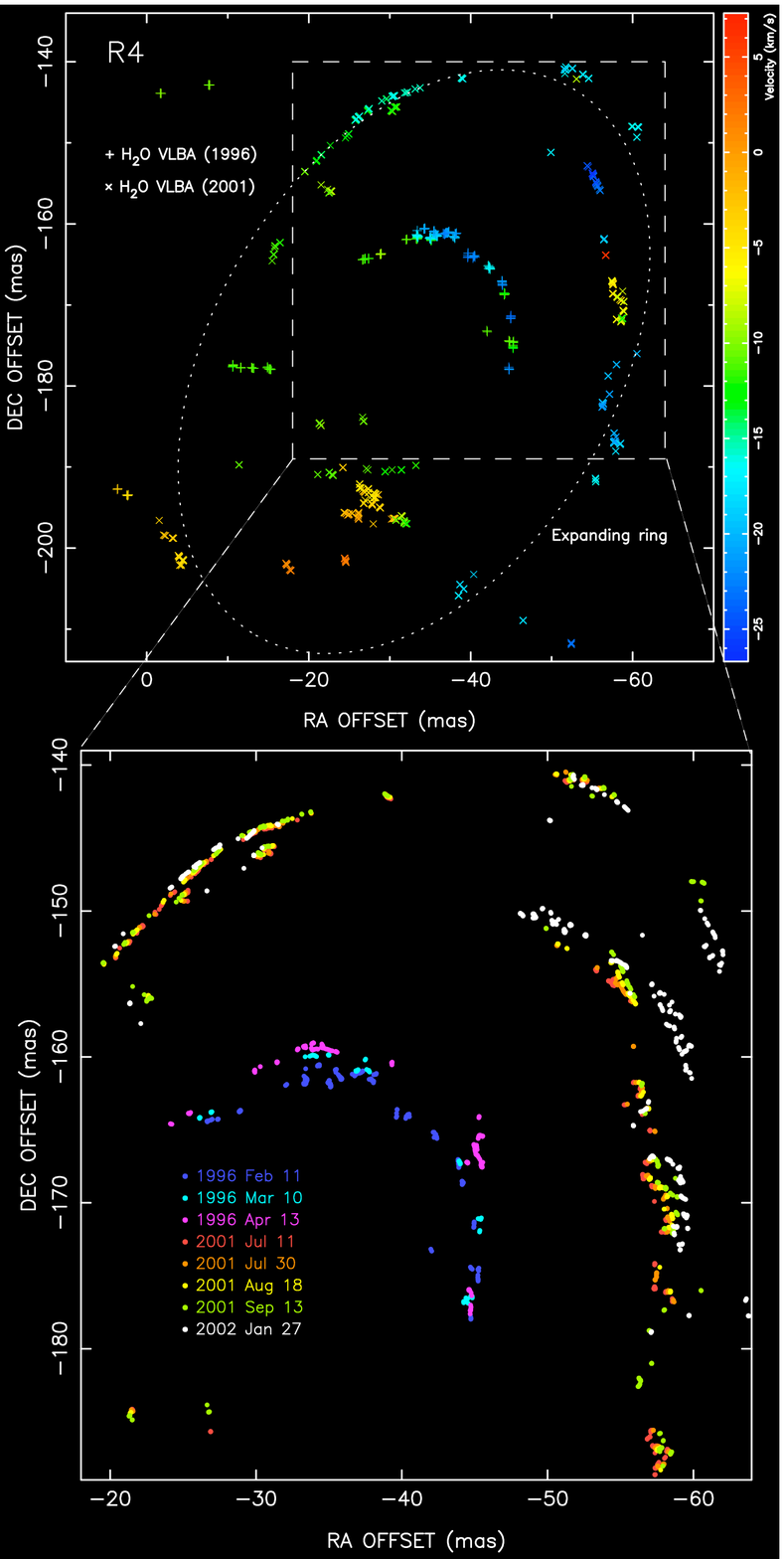}  
\end{center} \caption{(Published online) {\it
      Upper panel}: Water maser positions measured in sub-region R4 of
    Cepheus A (see also Figure~1).  Color code indicates the LSR radial velocity (km/s) of the masers. {\it Lower panel:} Zoom
    showing the evolution of the expanding motions in the sky for all
    the observed VLBA epochs. Color code indicates the epoch. (Figure from Torrelles et
    al. 2011).}
\end{figure}

Cepheus A is an active high-mass star-forming region, located at only
700 pc (Moscadelli et al. 2009), which 
contains a cluster of at least 16 sources within $\simeq 30''$
(Hughes \& Wouterloot 1984, Garay et al. 1996). Its
star-formation activity is dominated by the source HW2. The radio
continuum emission in HW2 seems to trace an ionized jet, powered by a
massive protostar ($\simeq 15-20$  M$_{\odot}$). Extremely broad millimeter wavelength hydrogen recombination lines were recently detected toward this source with terminal velocity $>$ 500 km~s$^-1$ for the ionized gas (Jim\'enez-Serra et al. 2011). This source
was the first identified case of a disk-protostar-jet system
in a massive YSO, at scales of $\simeq 1000$ AU
(Rodr\'{\i}guez et al. 1994, Patel et al. 2005, Curiel et al. 2006, Jim\'enez-Serra et al. 2007, Torrelles et al. 2007, Vlemmings et al. 2010). This suggests that this high-mass object formed by an accretion
mechanism, in a similar way to low-mass stars.

HW2 is associated with strong water maser emission, which has been
studied in detail with the Very Large Array (VLA, beam size $\simeq
80$ mas) and the Very Long
Baseline Array (VLBA, beam size $\simeq 0.5$ mas), for a total of 9 different
epochs (Torrelles et al. 2001a,b, 2011). Fig. 1 shows the location of
the water masers,  overlaid on the 1.3~cm continuum map of the
HW2 radio jet and the nearby HW3c and HW3d objects (located $\sim$
3$''$ south from HW2; see Chibueze et al. 2012 and Chibueze's talk in this conference for a detailed study of the distribution and kinematics of the water masers associated with HW3d). 
The VLBA data showed that individual maser
features tend to be organized both spatially and kinematically, forming 
linear microstructures of a few mas in size. The flattened appearance
of these water maser linear 
``microstructures'' and their proper motions indicate that they are
originated through shock excitation by outflows, as expected from theory
(e.g., Elitzur et al. 1992). These linear
microstructures
are the
building blocks of larger, coherent linear/arcuate structures of $\simeq
40-100$ mas. The identified structures in the neighborhood of HW2 are
in the regions labelled R1 to R8 in Figure~1. 

In particular, the masers in the R4 subregion trace a section of a nearly
elliptical ring of $\simeq 70$ mas size (50 AU), with expanding
motions of $\sim$ 15-30~km~s$^{-1}$ (Figure 2), from which an
extremely short dynamical age of 4-8 year can be derived. This
expansion must be driven by a yet undetected probably massive YSO (to explain the high luminosity of the water masers), located at the
geometrical center of the ring, $\sim 130$ AU from HW2. 
A detailed inspection of Fig. 2 also
shows some internal structure within the ring in individual epochs, 
with several
``shells'' that may have been created by multiple successive
ejections. 
Future sensitive cm and (sub)mm observations may help us to
identify the massive object powering this structure.

Another remarkable subregion is R5, located $\sim 6''$ south of
HW2. In our 1996 observations, the masers in this region traced part
of a nearly perfect circle (to an accuracy of 1/1000), and expanding at 9
km~s$^{-1}$. We interpreted this as
a short-lived (dynamical age $\simeq 30$ yr) 
spherical wind, powered by a high-mass object
at its center
(later detected in radio continuum observations by Curiel et
al. 2002). Only five years after its discovery, 
this maser structure has already
lost its spherical symmetry, probably due to its interaction with the
surrounding interstellar medium.
The discovery of a nearly spherical ejection is especially
relevant, since this phenomenon is difficult to explain in current models of
star-formation, which assume that mass loss in YSOs is produced by
the transformation of rotational energy in the disk into collimated
outflows via magnetohydrodynamic mechanisms. A few other cases of
nearly-isotropic ejections have been reported later in other
star-forming regions (e.g., W75N, see \S 3). This indicates that
uncollimated, episodic ejections may occur during the earliest stages
of evolution of massive YSOs. The mechanism powering those ejections is
still unclear. 

\begin{figure}[!ht]
\begin{center} 
\includegraphics[scale=.44]{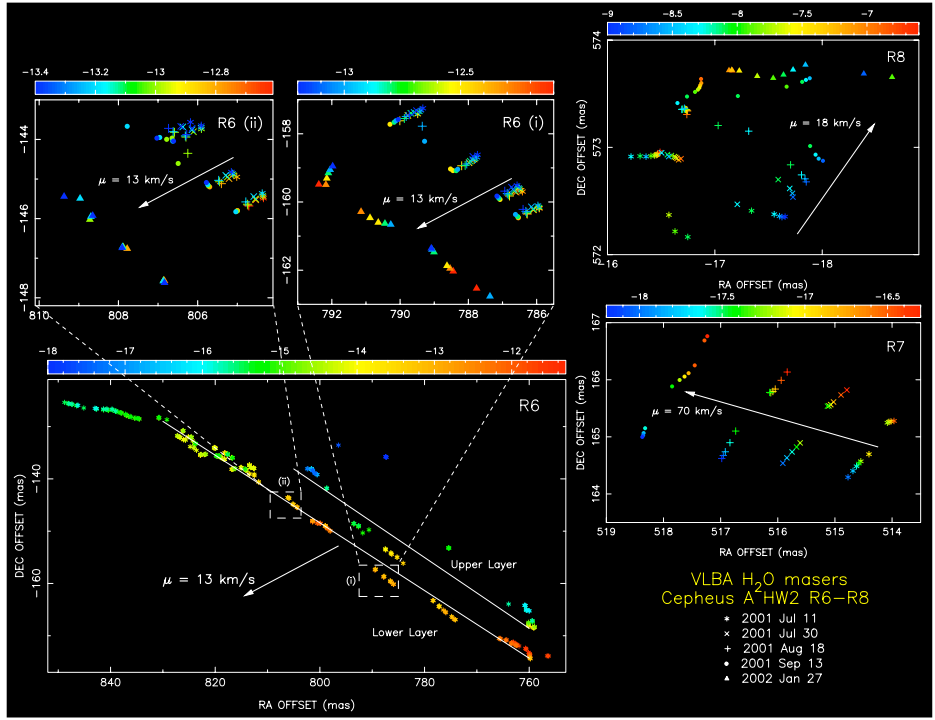} 
\end{center} 
\caption{(Published online) Positions
    and proper motions of the H$_2$  masers in sub-regions R6,
    R7, and R8 of Cepheus A (see also Figure 1). Color code indicates the LSR radial velocity
    (km~s$^{-1}$) of the masers. (Figure from Torrelles et
    al. 2011).} 
\end{figure} 

While the maser structures in R4 and R5 are believed to be associated
with different sources, the remaining ones (subregions R1 to R3, and
R6 to R8) could trace an outflow powered by HW2. Significant
differences in the magnitude and direction of the proper motions in
these structures (Fig 3.), suggest that we can be witnessing the
simultaneous presence of a collimated jet and a wide-angle outflow. In
this scenario (Fig. 4), R6 (moving to the southeast at $\sim 13$ km s$^{-1}$)
and R8 (to the northeast at $\sim 18$ km s$^{-1}$), would be tracing shock
fronts at the walls of expanding cavities created by a wide-angle outflow
from HW2, with an opening angle of $\simeq 100^\circ$. On the other
hand, there is clearly a collimated jet traced by the radio continuum
emission. The R7 masers, moving closer 
to the axis of this jet ($\simeq 30^\circ$ from the axis), 
show expansion velocities of $\sim 70$ km s$^{-1}$, which are intermediate
between those observed in R6/R8 ( $\sim 13-18$ km s$^{-1}$)
and  in the jet ($\sim 500$ km s$^{-1}$). The R1-R3 masers, located on
the opposite side from the central source, moving at $\sim$
5~km~s$^{-1}$ and with a difference in PA of  $\simeq$ --80$^{\circ}$
with respect to the radio jet direction, would represent the
corresponding  shocked walls of the southwestern cavities created by
the wide-angle outflow.

The simultaneous presence of a collimated and a wide-angle outflow has
been previously found in low-mass YSOs (e.g., see Velusamy et
al. 2011 and references therein). Different theoretical models have
been proposed to explain these two kind of outflows in low-mass YSOs
(e.g., ``X-wind'', ``Disk-wind'' models; see  Machida et
al. 2008). The evidence that they are also found in a high-mass
object, further reinforces that similar processes could be at work in
both low- and high-mass star formation. Moreover, these
observations in HW2 should provide important observational constraints
for future models trying to reproduce the presence of outflows with
different opening angles in high-mass YSOs.

\begin{figure}[!ht]
\begin{center} 
\includegraphics[scale=0.6]{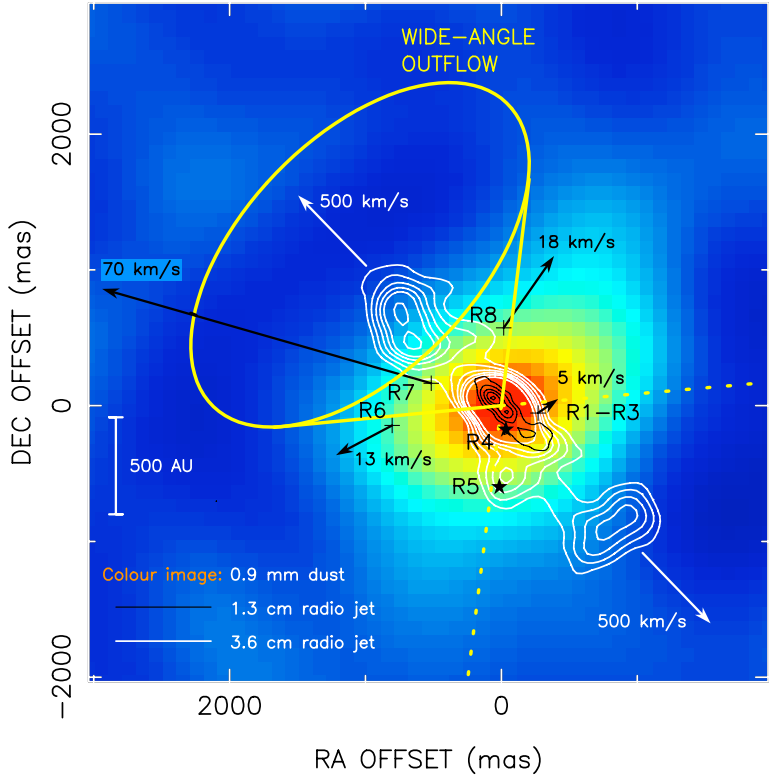}
\end{center}  
\caption{(Published online) Wide-angle
    outflow and jet in Cepheus A HW2. The radio jet (opening angle of
    $\sim$ 18$^{\circ}$) exhibits ejections in opposite directions,
    moving away at $\sim$ 500~km~s$^{-1}$ from the central source,
    and is surrounded by a dust/molecular disk (Patel et al. 2005). R6, R8, and R1-3 trace emission fronts from
    the shocked walls of expanding cavities, created by the wide-angle
    wind of HW2 (opening angle of $\sim$ 100$^{\circ}$). The R7
    masers, with motions along an axis at an angle of $\sim$
    30$^{\circ}$ with respect to the radio jet axis,  are excited
    inside the cavity by the wide-angle wind. They exhibit higher
    velocity than R6, R8, and R1-3 (which are located at the expanding
    cavity walls) but lower than the velocity of the jet. The R6, R7,
    and R8 masers (observed towards the blue-shifted lobe of the 1
    arcmin, large-scale bipolar molecular outflow; G\'omez et
    al. 1999) are blue-shifted with respect to the systemic velocity
    of the circumstellar disk, while R1-3 (observed towards the
    red-shifted lobe of the large-scale molecular outflow) are
    red-shifted. The position of the two massive YSOs required to
    excite the R4 and R5 maser structures are indicated by star
    symbols (see text). The star associated with R4 is not yet
    detected. (Figure and caption adapted from Torrelles et
    al. 2011, 2012).}
\end{figure} 


\section{W75N}

\begin{figure}[!ht]
\begin{center} 
\includegraphics[scale=.58]{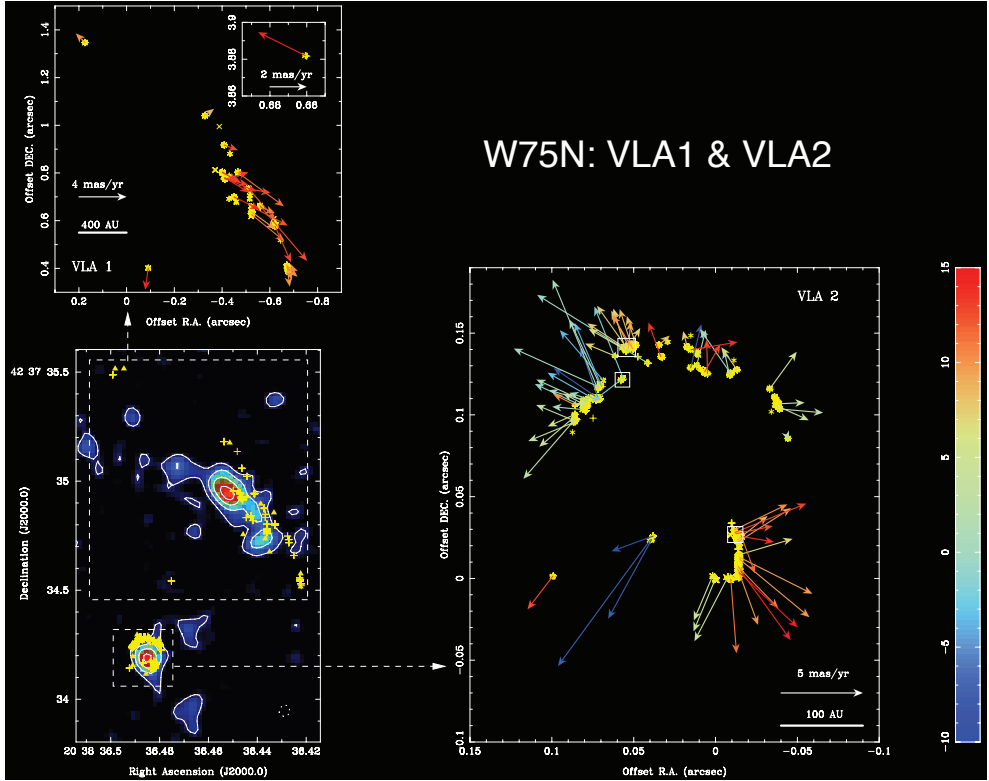} 
\end{center} 
\caption{(Published online) {\it
      Bottom left:} 1.3~cm continuum contour map of
    VLA 1 and VLA 2. The positions of the H$_2$O masers detected
    with the VLA (triangles) and VLBA (plus symbols) are
    indicated. {\it Top left and bottom right}: Proper motions (arrows) of the
    H$_2$O masers in VLA 1 and VLA 2. Color code
    indicates the LSR velocity of the masers in km~s$^{-1}$. (Figure
    from Torrelles et al. 2003, 2012).} 
\end{figure} 

We have focused our studies in three YSOs (VLA 1, VLA 2, VLA 3) in this
region of high-mass star formation (see, e.g., Hunter et al. 1994, Torrelles et al. 1997, Persi et al. 2006, and Carrasco-Gonz\'alez et al. 2010b).  Despite being located within a
region of only $1.5''$ (3000 AU at a distance of 2 kpc), they are
believed to be in different evolutionary stages. In particular, VLA 1
and VLA 2 show distinctive mass-loss characteristics (see Figure 5): while in VLA1 the water
masers observed with the VLBA, and the radio continuum emission trace
a collimated jet of $\sim 2000$ AU size, the radio continuum
emission in VLA 2 is compact, and the water masers seem to expand
with no preferential direction, tracing a shell-like outflow of $\sim
160$ AU radius, moving at
$\sim 30$ km~s$^{-1}$ (dynamical age of $\simeq 13$ yr). Given the
close proximity between these sources (projected distance 1400 AU),
they are likely to share a common molecular gas environment. This led
Torrelles et al. (2003) to 
suggest that their different degree of outflow collimation is
probably not due to ambient conditions, but to the different
evolutionary stages of these sources, with VLA2 being younger. 
In this scenario, it is expected that the VLA 2
outflow would become more collimated as the source evolves. In fact, 
recent VLBA water maser observations (Surcis et al. 2011) indeed
suggest that a jet is being formed in VLA 2.

The case of VLA 2 is reminiscent of that of the R5 subregion in
Cepheus A, suggesting the presence of non-collimated outflows in very
early stages of high-mass YSOs. It is still unclear whether 
the observed 
non-collimated outflows are non-standard phenomena in particular types
of sources, or if all 
massive YSOs undergo short-lived pulsed ejection phases with these type of processes. 
Nevertheless, their presence poses new challenges in our
knowledge of the earliest stages of stellar evolution.

\begin{acknowledgment}
GA, RE, JFG, and JMT acknowledge support from MICINN (Spain) AYA2008-06189-C03 and AYA2011-30228-C03 grants, co-funded with FEDER funds. SC acknowledges support from CONACyT (Mexico) 60581 and 168251 grants.
\end{acknowledgment}


\begin{thebibliography}{}

\bibitem []{}{Anglada, G. 1996, ASPC, 93, 3}

\bibitem []{}{Carrasco-Gonz\'alez, C., Rodr\'{\i}guez, L. F.,  Anglada, G.,  Mart\'{\i}, J., Torrelles, J. M., Osorio, M. 2010a, Science, 330, 1209}

\bibitem []{}{Carrasco-Gonz\'alez, C. et al. 2011, RMxAC, 40, 229}

\bibitem []{}{Carrasco-Gonz\'alez, C. et al. 2010b, AJ, 139, 2433}

\bibitem []{}{Chibueze, J. M. et al. 2012, ApJ, in press}

\bibitem []{}{Curiel, S. et al. 2006, ApJ, 638, 878}

\bibitem []{}{Curiel, S. et al. 2002, ApJ, 564, L35}

\bibitem []{}{Davies, B., Lumsden, S. L., Hoare, M. G., Oudmaijer, R. D., de Wit, W-J. 2010, MNRAS, 402, 1504}

\bibitem []{}{Elitzur, M., Hollenbach, D. J., McKee, C. F. 1992, ApJ, 394, 221}

\bibitem []{}{Fern\'andez-L\'opez, M., Girart, J. M., Curiel, S., G\'omez, Y., Ho, P. T. P., Patel, N. 2011, AJ, 142, 97}


\bibitem []{}{Garay, G., Lizano, S. 1999, PASP, 111, 1049}

\bibitem []{}{Garay, G., Ram\'{\i}rez, S., Rodr\'{\i}guez, L.~F., Curiel, 
S., Torrelles, J.~M. 1996, 1996, ApJ, 459, 193}

\bibitem []{}{Goddi, C., Moscadelli, L., Sanna, A. 2011, A\&A, 535, L8}

\bibitem []{}{G\'omez, J. F. et al. 1999, ApJ, 514, 287}

\bibitem []{}{Hughes, V. A., Wouterloot, J. G. A. 1984, ApJ, 276, 204}

\bibitem []{}{Hunter, T. R., Taylor, G. B., Felli, M., Tofani, G. 1994, A\&A, 284, 215}


\bibitem []{}{Jim\'enez-Serra, I. et al. 2007, ApJ, 661, L187}

\bibitem []{}{Jim\'enez-Serra, I. et al. 2011, ApJ, 732, L27}


\bibitem []{}{Machida, M. N., Inutsuka, S.-i., Matsumoto, T. 2008, ApJ, 676, 1088}

\bibitem []{}{Moscadelli, L. et al. 2009, ApJ, 693, 406}

\bibitem []{}{Motogi, K. et al. 2008, MNRAS, 390, 523}

\bibitem []{}{Patel, N. A., Curiel, S., Sridharan, T. K., 
Zhang, Q., Hunter, T. R., Ho, P.T.P., Torrelles, J. M., Moran,
J. M., G\'omez, J. F. G., Anglada, G. 2005, Nature, 437, 109}

\bibitem []{}{Persi, P., Tapia, M., Smith, H. A. 2006, A\&A, 445, 971}

\bibitem []{}{Reid, M. J., Moran, J. M. 1981, ARAA, 19, 231}

\bibitem []{}{Rodr\'{\i}guez, L.~F., Garay, G., Curiel, S., Ram\'{\i}rez, S., Torrelles, J.~M., G\'{o}mez, Y., Vel\'{a}zquez, A. 1994, ApJ, 430, L65}

\bibitem []{}{Sanna, A., Reid, M. J., Carrasco-Gonz\'alez, C., Menten, K. M., Brunthaler, A., Moscadelli, L., Rygl, K. L. J. 2012, ApJ, 745, 191}

\bibitem []{}{Surcis, G.,Vlemmings, W. H. T., Curiel, S., Hutawarakorn Kramer, B., Torrelles, J. M., Sarma, A. P. 2011, A\&A, 527, 48}

\bibitem []{}{Torrelles, J. M., G\'omez, J. F., Rodr\'{\i}guez, L. F., Ho, P. T. P., Curiel, S., V\'azquez, R. 1997, ApJ, 489, 744}

\bibitem []{}{Torrelles, J. M., Patel, N., Anglada
G., G\'omez, J. F., Ho, P. T. P., Cant\'o, J., Curiel, S., 
Lara, L., Alberdi, A., Garay, G., Rodr\'{\i}guez, L. F. 2003, ApJ, 598, L115}

\bibitem []{}{Torrelles, J. M., Patel, N. A., Curiel, S., Estalella, R., G\'omez, J. F., Rodr\'{\i}guez, L. F., Cant\'o, J., Anglada, G., Vlemmings, W., Garay, G., Raga, A. C., Ho, P. T. P. 2011, MNRAS, 410, 627}

\bibitem []{}{Torrelles, J. M., Patel, N. A., Curiel, S., Ho, P. T. P., Garay, G., Rodr\'{\i}guez, L. F. 2007, ApJ, 666, L37}

\bibitem []{}{Torrelles, J. M., Patel, N., G\'omez, J. F., 
Ho, P. T. P., Rodr\'{\i}guez, L. F., Anglada, G., Garay, G., Greenhill, L.,
Curiel, S., Cant\'o, J. 2001a, Nature, 411, 277}

\bibitem []{}{Torrelles, J. M., Patel, N., G\'omez, J. F., 
Ho, P. T. P., Rodr\'{\i}guez, L. F., Anglada, G., Garay, G., Greenhill, L.,
Curiel, S., Cant\'o, J. 2001b, ApJ, 560, 853}

\bibitem []{}{Torrelles, J. M., Patel, N., Curiel, S., G\'omez, J. F., 
Anglada, G., Estalella, R. 2012,  Bolet\'{\i}n Asoc. Argentina de Astronom\'{\i}a, Eds. J.J. Clari\`a, A.E. Piatti, R. Barb\'a, P. Benaglia and F. Bareilles, No. 54, in press}

\bibitem []{}{Trinidad, M. et al. 2003, ApJ, 589, 386}

\bibitem []{}{Velusamy, T., Langer, W. D., Kumar, M. S. N., Grave, J. M. C. 2011, ApJ, 741, 60}

\bibitem []{}{Vlemmings, W. H. T. et al. 2010, MNRAS, 404, 134}

\bibitem []{}{Zapata, L. A. et al. 2009, ApJ, 698, 1422}

\end{thebibliography}
\end{document}